\documentclass[preprint,showpacs,preprintnumbers,amsmath,amssymb]{revtex4}
\usepackage{graphics,epsfig,subfigure}
\usepackage{color}

\begin{document}
\renewcommand{\baselinestretch}{1.3}
\newcommand\beq{\begin{equation}}
\newcommand\eeq{\end{equation}}
\newcommand\beqn{\begin{eqnarray}}
\newcommand\eeqn{\end{eqnarray}}
\newcommand\nn{\nonumber}
\newcommand\fc{\frac}
\newcommand\lt{\left}
\newcommand\rt{\right}
\newcommand\pt{\partial}

\title{Domain Wall Brane in Eddington-Inspired Born-Infeld Gravity}
\author{Yu-Xiao Liu\footnote{liuyx@lzu.edu.cn},
        Ke Yang\footnote{yangke09@lzu.edu.cn, corresponding author},
        Heng Guo\footnote{guoh2009@lzu.edu.cn}
        and Yuan Zhong\footnote{zhongy2009@lzu.edu.cn}}.
 \affiliation{Institute of Theoretical Physics, Lanzhou University, Lanzhou 730000,
             China}

\begin{abstract}
Recently, inspired by Eddington's theory, an alternative gravity called Eddington-inspired Born-Infeld  gravity was proposed by Ba$\tilde{\text{n}}$ados and Ferreira. It is equivalent to Einstein's general relativity in vacuum, but deviates from it when matter is included. Interestingly, it seems that the cosmological singularities are prevented in this theory. Based on the new theory, we investigate a thick brane model with a scalar field presenting in the five-dimensional background. A domain wall solution is obtained, and further, we find that at low energy the four-dimensional Einstein gravity is recovered on the brane. Moreover, the stability of gravitational perturbations is ensured in this model.
\end{abstract}

% \Keywords{ }


%\pacs{04.50.-h, 11.27.+d }

%04.50.-h Higher-dimensional gravity and other theories of gravity
%         (see also 11.25.Mj Compactification and four-dimensional models, 11.25.Uv D branes)

%04.50.Kd Modified theories of gravity

% 04.50.+h Gravity in more than four dimensions, Kaluza-Klein theory,
           % unified field theories, alternative theories of gravity
           %(see also 11.25.M Compactification and four-dimensional models), dilaton gravity
% 11.27.+d Extended classical solutions; cosmic strings,
           %domain walls, texture (see 98.80.C in cosmology)

\maketitle

\section{Introduction}

As the mainstay of gravitational theory, Einstein's general relativity (GR) provides precise descriptions to a variety of phenomena in our Universe. But, it is also well known that GR suffers various troublesome theoretical problems, such as the dark matter problem \cite{Bertone2005}, dark energy problem \cite{Li2011a}, and the singularity problem \cite{Hawking1970}. Thus, alternative theories are helpful to enrich our knowledge about the gravity and may provide us new approaches to overcome these problems.

A purely affine gravitational theory was introduced by Eddington in 1924 \cite{Eddington1924,Schrodinger1950}. The theory is totally equivalent to the GR with cosmological constant presented. However, Eddington's theory is incomplete because matter is not included. Recently, inspired by Eddington's theory, a new intriguing theory called Eddington-inspired Born-Infeld (EiBI) theory was put forward by Ba$\tilde{\text{n}}$ados and Ferreira \cite{Banados2010}. The authors completed Eddington's theory with matter included in a conventional way. Instead of insisting on a purely affine action, they worked in a Palatini formulation, i.e., the metric and the connection are regarded as independent fields, and switched the structure to a Born-Infeld-like one \cite{Born1934}. The EiBI theory reproduces GR appending some high-order terms of Ricci tensor for a small matter density, and approximates to Eddington's for a large one. Furthermore, it can be shown that this theory is completely equivalent to GR with matter fields absent, but deviates from GR in the presence of matter. As the most attractive feature, cosmological singularities seem to be prevented in this theory \cite{Banados2010,Pani2011}. The relevant cosmological and astrophysical issues were considered in \cite{Pani2011,Casanellas2012,Avelino2012,Pani2012}, the singularity features of EiBI theory with perfect fluids as the matter were included in \cite{Delsate2012}, the tensor perturbations of a homogeneous and isotropic space-time were discussed in \cite{Escamilla-Rivera2012}, and a nonsingular bouncing universe considered as a new solution to the instability problem of tensor perturbations was proposed in \cite{Avelino2012a}, recently.

On the other hand, in order to unify electromagnetism and gravity, Kaluza and Klein (KK) proposed a five-dimensional Einstein's theory with a circle as the extra spatial dimension in the 1920s \cite{Kaluza1921,Klein1926}. The KK theory opens the way to describe the particle interactions in higher-dimensional space-time. Because of some problems in this theory, KK's pioneering idea had not draw enough attention until the late 1970s and 1980s with the developments of superstring theories. In order to address the problem of why physical properties of the observed four-dimensional space-time are totally different from the extra dimensions, the brane-world scenario as one of the possible mechanisms was suggested. The prototype idea of brane world was proposed during the early 1980s \cite{Akama1982,Rubakov1983} and made great progress after the  Arkani-Hamed-Dimopoulos-Dvali model \cite{Arkani-Hamed1998,Antoniadis1998} and Randall-Sundrum model \cite{Randall1999,Randall1999a} proposed in the late 1990s. It suggests that the standard model particles are trapped on a four-dimensional hypersurface (called brane)
embedded in a higher-dimensional space-time (called bulk). There has been increasing interest during recent years in brane-world scenario \cite{Goldberger1999a,Gremm2000a,Gremm2000,DeWolfe2000,Csaki2000,Gherghetta2000,Arkani-Hamed2001a,Campos2002,Kobayashi2002,Wang2002,Charmousis2003,Bazeia2004a,Liu2007,
Dzhunushaliev2009,Dzhunushaliev2010a,Liu2011}.
This is because it provides us new perspectives to solve some disturbing problems in high-energy physics, such as the gauge hierarchy problem and the cosmological constant problem \cite{Arkani-Hamed1998,Antoniadis1998,Randall1999,Randall1999a}. And, it may also open up new horizons to understand our Universe, see, e.g., Refs. \cite{Rubakov2001,Csaki2004,Cheng2010} for introduction.

Based on the gravity coupled to the background scalar fields in multidimensional space-time, the brane configuration is determined by the gravity theory, the scalar fields, and the ways of scalar-gravity coupling. In this paper, we are interested in the brane-world scenario based on this new gravitational theory, the EiBI theory. A background scalar field is included in the five-dimensional bulk to generate the smooth thick brane configuration.
We shall show that the domain wall solution is supported by the theory, and further, the linear tensor perturbations are stable and four-dimensional Einstein gravity is recovered as the low-energy effective theory on the brane.

The paper is organized as follows: In Sec. \ref{The_EiBI_Theory}, we give a brief introduction to the EiBI theory. In Sec. \ref{The_Model}, we put forward a thick brane model and solve the theory to get a domain wall solution. In Sec. \ref{Gravitational_Fluctuations}, gravitational fluctuations are considered. Finally, conclusions and discussion are presented.

\section{The $n-$dimensional Eddington-inspired Born-Infeld Theory}\label{The_EiBI_Theory}

Following Refs. \cite{Banados2010,Vollick2004}, we consider the EiBI theory in $n$-dimensional space-time and the action is proposed as
\beqn
S(g,\Gamma,\Phi)=\fc{1}{\kappa b}\int d^{n}x\lt[\sqrt{-|g_{MN}+bR_{MN}(\Gamma)|}\rt.
        -\lt. \lambda\sqrt{-|g_{MN}|}\rt]+S_{M}(g,\Phi), \label{EiBI_action}
\eeqn
where $\kappa=8\pi G_5$, $b$ is a constant with inverse dimensions to that of cosmological constant, and $R_{MN}(\Gamma)$ represents the symmetric part of the Ricci tensor built with the connection $\Gamma$. $S_{M}(g,\Phi)$ is the action of matter fields coupled to the metric only. The dimensionless parameter $\lambda$ must be different from zero, inasmuch as  when matter fields are absent, the metric variation yields $\sqrt{-|g_{PQ}+b R_{PQ}|}[(g_{PQ}+b R_{PQ})^{-1}]^{MN}=0$ , and this makes no sense.

The equations of motion for this theory are obtained by varying the action (\ref{EiBI_action}) with respect to the metric field $g$ and the connection field $\Gamma$, respectively. The variation of the action with respect to the metric simply gives
\beq
\fc{\sqrt{-|g_{PQ}+bR_{PQ}|}}{\sqrt{-|g_{PQ}|}}[(g_{PQ}+bR_{PQ})^{-1}]^{MN}-\lambda g^{MN}
=-\kappa b T^{MN},\label{EOM1}
\eeq
where the energy-momentum tensor is defined as $T^{MN}=\fc{2}{\sqrt{-|g_{PQ}|}}\fc{\delta L_{M}(g,\Phi)}{\delta g_{MN}}$ with indices raised by the metric $g_{MN}$.

The variation with respect to the connection can be simplified by introducing an auxiliary metric
\beq
q_{MN}=g_{MN}+bR_{MN},\label{EOM2}
\eeq
and hence the variation leads to $q_{MN;K}=0$, where the semicolon is the covariant derivative with respect to the connection $\Gamma$. This means that the auxiliary metric $q_{MN}$ is compatible with the connection $\Gamma$, i.e., ${\Gamma^{K}_{MN}}=\fc{1}{2}q^{KL}(q_{LM,N}+q_{LN,M}-q_{MN,L})$ is the Christoffel symbol of the auxiliary metric. Then by combining (\ref{EOM1}) and (\ref{EOM2}), one has
\beq
\sqrt{-|q_{PQ}|}~q^{MN}=
\lambda\sqrt{-|g_{PQ}|}~g^{MN}-b\kappa\sqrt{-|g_{PQ}|}~T^{MN}, \label{EOM3}
\eeq
where $q^{MN}$ is the inverse of $q_{MN}$.

The Eqs. (\ref{EOM2}) and (\ref{EOM3}) and  matter field equations form a complete set of equations of the theory.

For a large value of $bR_{MN}$, the EiBI action (\ref{EiBI_action}) apparently approximates to the Eddington's, while for a small value of $bR_{MN}$, by expanding the first term $\sqrt{-|g_{MN}+bR_{MN}|}$ of the EiBI action to second order in $b$, one has
\beqn
S=\fc{1}{2\kappa}\int d^{n}x\sqrt{-|g_{MN}|}\lt[R-2\Lambda_{\text{eff}}+\fc{b}{4}RR \rt.
-\lt.\fc{b}{2}{R^M}_N{R^N}_M+\mathcal{O}(b^2)\rt] +S_{M}(g,\Phi),
\label{Smal_Appro_Action}
\eeqn
where $R=g^{MN}R_{MN}(q)$ and $\Lambda_{\text{eff}}\equiv(\lambda-1)/b$. Thus, Eqs. (\ref{EOM2}) and (\ref{Smal_Appro_Action}) clearly show that the EiBI action reproduces the Einstein-Hilbert action with cosmological constant $\Lambda_{\text{eff}}$ in lowest-order approximation. Further, by varying this approximate action with respect to the metric $g_{MN}$, one has the modified Einstein equations
\beqn
R_{MN}\!&\!=\!&\!\fc{2}{n-2}\Lambda_{\text{eff}}~g_{MN}+\kappa\lt[T_{MN}-\fc{1}{n-2}Tg_{MN}\rt]
+b\kappa^2\lt[S_{MN}-\fc{1}{2(n-2)}Sg_{MN}\rt] \nn\\
\!&\!-\!&\!\fc{n-4}{n-2}b\Lambda_{\text{eff}}
          \times\lt[\fc{1}{n-2}\Lambda_{\text{eff}}~g_{MN}+\kappa(T_{MN}-\fc{1}{n-2}Tg_{MN})\rt],~~\label{Correction_EEq}
\eeqn
where $S_{MN}={T^{K}}_{M}T_{KN}-\fc{1}{n-2}TT_{MN}$. When $n=4$, the modified Einstein equation (\ref{Correction_EEq}) will degenerate into the standard Einstein equations with matter absent \cite{Banados2010}. However, it is not the same case when $n=5$, for there is still an additional correction associated with cosmological constant. But, this is not in conflict with the conclusion that the EiBI action (\ref{EiBI_action}) is equivalent to the Einstein-Hilbert action in vacuum, since when matter is absent, Eq. (\ref{EOM1}) simply implies that
\beqn
R_{MN}=\fc{\lambda^{\fc{2}{n-2}}-1}{b}g_{MN}. \label{Vac_Eq1}
\eeqn
Then, we substitute this relation into the field equation (\ref{EOM2}), and it just gives us
\beqn
q_{MN}=\lambda^{\fc{2}{n-2}}g_{MN}. \label{Vac_Eq2}
\eeqn
It means that the metric $g_{MN}$ is also compatible with the connection $\Gamma$. Thus, Eqs. (\ref{Vac_Eq1}) and (\ref{Vac_Eq2}) indicate that the EiBI action is equivalent to the Einstein-Hilbert action with the cosmological constant $\Lambda_\text{G}=(n/2-1)(\lambda^{\fc{2}{n-2}}-1)/b$ when matter is absent. Without loss of generality, here we assume that the parameter $b$ is positive, thus when $\lambda<1$, $\Lambda_\text{G}<0$ represents an anti-de Sitter (AdS) vacuum; when $\lambda>1$, $\Lambda_\text{G}>0$ represents a de Sitter (dS) vacuum;  and when $\lambda=1$, $\Lambda_\text{G}=0$ represents a Minkowski vacuum. Furthermore, when the parameter $\lambda\neq1$, substituting (\ref{Vac_Eq1}) back into the EiBI action with $S_M=0$ just gives the Eddington action \cite{Schrodinger1950,Banados2010}
\beq
S(\Gamma)=\fc{1}{\kappa\tilde{b}}\int{d^nx\sqrt{-|\tilde{b}R_{MN}(\Gamma)|}},
\label{Edd_Action}
\eeq
where the parameter $\tilde{b}{\equiv}b/(1-\lambda^{\fc{2}{2-n}})$.

\section{The Model}\label{The_Model}

In this section, based on the five-dimensional EiBI theory, we consider a brane-world model with a scalar field existing in the background as the ``material" to construct the brane configuration. The ansatz for the most general metric which preserves four-dimensional Poincar$\acute{\text{e}}$ invariance is \cite{Randall1999a}
\beq
ds^2=a^2(y)\eta_{\mu\nu}dx^{\mu}dx^{\nu}+dy^2, \label{RS_metric}
\eeq
where $a^2(y)$ is the warp factor and $y$ denotes the physical coordinate of extra dimension. The EiBI theory without matter is fully equivalent to Einstein-Hilbert action with the cosmological constant $\Lambda_\text{G}$. While when matter such as here a scalar field existed in the background space-time, the EiBI theory will be distinct from the Einstein's relativity. The Lagrangian of a scalar field is given by
\beq
L_{M}=-\sqrt{-|g_{PQ}|}\lt[\fc{1}{2}\pt^{K}\phi\pt_{K}\phi+V(\phi)\rt],
\eeq
where $V(\phi)$ is the scalar potential. We assume that the scalar field refers to the extra dimension only, i.e., $\phi=\phi(y)$, to be consistent with the four-dimensional Poincar$\acute{\text{e}}$ invariance of the metric. Then, the scalar field equation is
\beq
\fc{1}{\sqrt{-|g_{PQ}|}}\pt^{K}\lt[\sqrt{-|g_{PQ}|}\pt_K\phi\rt]=\fc{\pt V}{\pt \phi}, \label{Scalar_EOM}
\eeq
and the contravariant energy-momentum tensor is given by
\beqn
T^{MN}=\pt^{M}\phi\pt^{N}\phi-\lt[\fc{1}{2}\pt^{K}\phi\pt_{K}\phi+V(\phi)\rt]g^{MN}.
\label{E_M_tensor}
\eeqn

In this case, the auxiliary metric can be assumed as $q_{MN}=(-u, u, u, u, v)$ with $u$ and $v$ the functions of $y$. From the Eqs. (\ref{EOM3}) and (\ref{E_M_tensor}), one finds
\beqn
uv^{\fc{1}{2}}&=& a^2\lt[\lambda+b\kappa\lt(\fc{1}{2}\phi'^2+V(\phi)\rt)\rt],\\\label{Eq1_for_qmetric}
u^2v^{-\fc{1}{2}}&=& a^4\lt[\lambda-b\kappa\lt(\fc{1}{2}\phi'^2-V(\phi)\rt)\rt], \label{Eq3_for_qmetric}
\eeqn
where the prime denotes the derivative with respect to the extra dimension $y$. Thus, we have
\begin{subequations}\label{Aux_Metric}
\beqn
u(y)\!&\!=\!&\!a^2\!\Big[\lambda+b\kappa\big(V+\fc{1}{2}\phi'^2\big)\Big]^{\fc{1}{3}}\Big[\lambda+b\kappa\big(V-\fc{1}{2}\phi'^2\big)\Big]^{\fc{1}{3}},~~~~~~~\\
v(y)\!&\!=\!&\!\Big[\lambda+b\kappa\big(V+\fc{1}{2}\phi'^2\big)\Big]^{\fc{4}{3}}\Big[\lambda+b\kappa\big(V-\fc{1}{2}\phi'^2\big)\Big]^{-\fc{2}{3}}.~~~~~~~
\eeqn
\end{subequations}
From the metric (\ref{RS_metric}), the equations of motion (\ref{EOM2}) and (\ref{Scalar_EOM}) are written explicitly as
\begin{subequations}\label{EOM_EP}
\beqn
&&u=a^2+b \fc{u u' v'-2 v(u'^2+u u'')}{4u v^2}, \label{EOM_EP_1}\\
&&v=1+b \fc{u u' v' +v(u'^2-2 u u'')}{u^2 v}, \label{EOM_EP_2}\\
&&4\fc{a'}{a}\phi'+\phi''=\fc{\pt V(\phi)}{\pt \phi}. \label{EOM_EP_3}
\eeqn
\end{subequations}

Since the form of the scalar potential $V(\phi)$ is not given, the system is not completely determined. Thus, the three equations are not totally independent and the three variables $a(y)$, $\phi(y)$, and $V(\phi)$ cannot be solved uniquely from the equations. With the expectation that the scalar is a kink solution (odd function), and the warp factor is a $Z_2$-symmetric function (even function), which peaks at the origin and
falls off along the extra dimension to make sure that the null signal takes an infinite amount of time to travel from $y=\pm\infty$ to $y=0$, we assume a simple restriction $\phi'(y)=K a^2(y)$ with $K$ a constant parameter. Then, the Eq. (\ref{EOM_EP_3}) can be easily solved as
\beq
V(y)=\fc{3}{2}K^2a(y)^4+V_0, \label{Scalar_potential}
\eeq
where the integral constant $V_0$ represents the scalar vacuum energy density. Thus, (\ref{Aux_Metric}) can be expressed as
$u(y)=a^2(\tilde\lambda+2b\kappa K^2a^4)^{\fc{1}{3}}(\tilde\lambda+b\kappa K^2a^4)^{\fc{1}{3}},
v(y)=(\tilde\lambda+2b\kappa K^2a^4)^{\fc{4}{3}}(\tilde\lambda+b\kappa K^2a^4)^{-\fc{2}{3}},$
where the parameter $\tilde\lambda=\lambda+b\kappa{V_0}$. Here, in order to simplify the calculation, we fix the  integral constant $V_0$ by setting $\tilde\lambda=0$, namely, we fix the scalar vacuum energy density as $V_0=-\fc{\lambda}{b\kappa}$. Then, the auxiliary metric is largely simplified as
\begin{subequations}\label{Aux_Metric_exp}
\beqn
u(y)\!&\!=\!&\!\alpha~a(y)^{\fc{14}{3}},\\
v(y)\!&\!=\!&\!2\alpha~a(y)^{\fc{8}{3}},
\eeqn
\end{subequations}
where the parameter $\alpha=(\sqrt{2} b\kappa K^2 )^{\fc{2}{3}}$. Now, it is easy to check that Eq. (\ref{EOM_EP}) supports the following solution:
\beqn
a(y)\!&\!=\!&\!\text{sech}^{\fc{3}{4}}({\fc{2}{\sqrt{21b}}y}),\label{Sol_Warp_factor}\\
\phi(y)\!&\!=\!&\!\pm\fc{7^{5/4}}{2\times3^{1/4}\kappa^{1/2}}\lt(i \text{E}(\fc{iy}{\sqrt{21 b}},2)+\text{sech}^{\fc{1}{2}}({\fc{2y}{\sqrt{21b}}})
\times\sinh({\fc{2y}{\sqrt{21b}}})\rt),\label{Sol_Scalar}
\eeqn
with the parameter $\alpha$ fixed as $\alpha=7/6$, namely,
\beq
K=\pm\big(\fc{7}{3}\big)^{\fc{3}{4}}\fc{1}{2\sqrt{b\kappa}}.
\eeq
%%%%%%%%%%%%%%%%%%%%%%%%%%%%%%%%%%%%%%%%%%%%%%%%%%%%%%%%%%%%%%%%
\begin{figure*}[htb]
\begin{center}
\subfigure[$\phi(y)$]  {\label{Scalar}
\includegraphics[width=7cm,height=5cm]{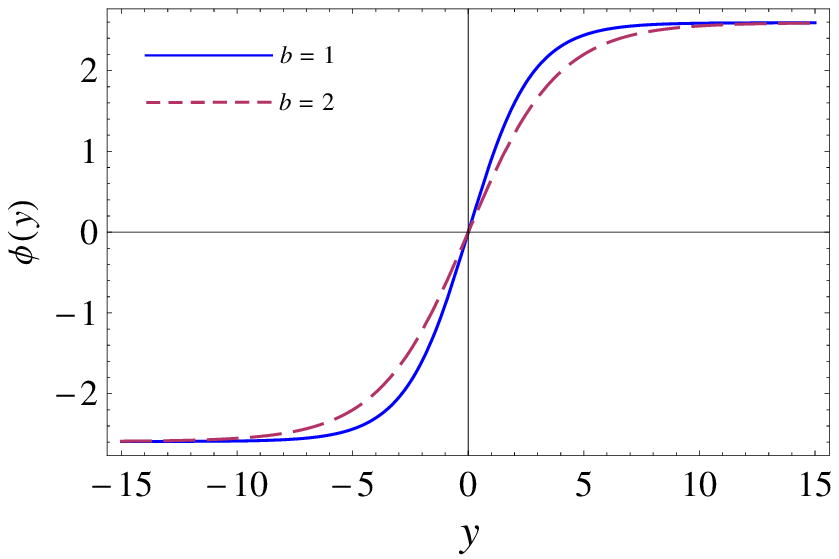}}
\subfigure[$\rho(y)$]  {\label{Fig_Energy_density}
\includegraphics[width=7cm,height=5cm]{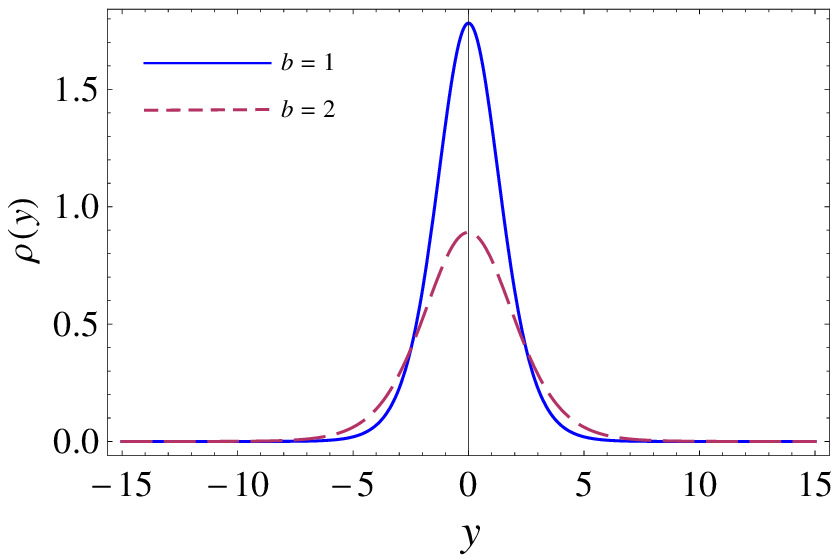}}
\end{center}
\caption{The shapes of the scalar $\phi(y)$ and the energy density $\rho(y)$. The
parameters are set to $ \kappa=1, \lambda=1$.}
\end{figure*}
%%%%%%%%%%%%%%%%%%%%%%%%%%%%%%%%%%%%%%%%%%%%%%%%%%%%%%%%%%%%%%%%%%
Here we have set the integral parameters to fix $a(0)=1$ and the function $\text{E}$ is an elliptic integral. We plot the branch of the scalar with respect to the positive $K$ in Fig. \ref{Scalar}, and the other branch is easily got by reflecting these curves along the horizontal axis. As $y\rightarrow\pm\infty$, the scalar $|\phi(y)|$ approaches a constant $v_0=i(\fc{7}{3})^{1/4}\fc{7}{4\sqrt{\kappa}}\lt(\sqrt{\fc{2}{\pi}}\Gamma^2(\fc{3}{4})-2\text{E}(2)\rt)\simeq{2.59}/{\sqrt{\kappa}}$.

Moreover, from Eq. (\ref{Scalar_potential}), the scalar potential $V(y)$ is given by
\beq
V(y)=\fc{7\sqrt{21}}{24b\kappa}\text{sech}^{3}({\fc{2y}{\sqrt{21b}}})-\fc{\lambda}{b\kappa}. \label{Scalar_Potential}
\eeq
With an elliptic integral existing in the expression of scalar, we fail to get an analytic expression of the scalar potential $V(\phi)$. Nevertheless, from Eq. (\ref{Scalar_potential}), we have $\fc{\pt V}{\pt \phi}=6Kaa'$ and $\fc{\pt^2 V}{\pt \phi^2}=6(\fc{a''}{a}+\fc{a'^2}{a^2})$, then we find that, as $\phi\rightarrow\pm v_0$, $\fc{\pt V}{\pt \phi}\rightarrow0$ and  $\fc{\pt^2 V}{\pt \phi^2}\rightarrow \fc{9b}{7}>0$. It means that the points $V(\pm v_0)$ are two minimums (vacua) of the potential. Thus, as we have expected, the scalar is indeed a kink solution with $\phi(\pm\infty)=\pm v_0$ corresponding to the two vacua of the potential. This brane solution depicts a domain wall configuration.

Further, the energy density is defined as $\rho(y)=T_{MN}w^{M}w^{N}-V_0$, where $w^M$ is the four-velocity of the static observer and here we have subtracted the contribution of the vacuum energy density $V_0$.  Thus, from Eq. (\ref{E_M_tensor}), we have
\beqn
\rho(y)=-T^0_0+\fc{\lambda}{b\kappa}=\fc{7\sqrt{21}}{18b\kappa}\text{sech}^{3}({\fc{2y}{\sqrt{21b}}}).
\eeqn
The energy density is localized in the origin of 5th dimension as shown in Fig. \ref{Fig_Energy_density}, and it does not dissipate with time. The brane thickness can be defined as the full width of the peak at half maximum value and it is completely decided by the parameter $b$. For a small value of $b$, the brane thickness can be hidden from our low-energy observation on the brane.

Moreover, with the warp factor and auxiliary metric, the Ricci scalar curvature is given by
\beq
R=g^{MN}R_{MN}=\fc{1}{b}\lt[2-7\tanh^2(\fc{2}{\sqrt{21b}})y\rt].
\eeq
When $y$ approaches infinity, the scalar curvature $R\rightarrow-5/b<0$. It means that the bulk is an asymptotically AdS space-time. It is also consistent with the brane configuration that matter mainly distributes on the brane with AdS vacuum left far away from it.

\section{Gravitational Fluctuations}\label{Gravitational_Fluctuations}

Here, we are interested in the tensor fluctuations which involve the spin-2 gravitons in our model, thus we impose the axial gauge $h_{\mu5}=h_{55}=0$ to remove the scalar mode and vector mode, and the perturbed metric is simply given by \cite{Randall1999}
\beqn
d\hat{s}^2=\hat{g}_{MN}(x,y)dx^Mdx^N
=a^2(y)[\eta_{\mu\nu}+h_{\mu\nu}(x,y)]dx^{\mu}dx^{\nu}+dy^2,
\label{Fluc_Metric_UC}
\eeqn
where $h_{\mu\nu}$ represent tensor fluctuations about background space-time. Then, with the relation (\ref{Aux_Metric_exp}), the perturbed auxiliary metric can be assumed as
\beqn
d\hat{s}'^2\!&\!=\!&\!\hat{q}_{MN}(x,y)dx^Mdx^N
=[q_{MN}+\xi_{MN}(x,y)]dx^Mdx^N\nn\\
\!&\!=\!&\!\alpha~ a^{\fc{8}{3}}(y)\lt[a^2(y)(\eta_{\mu\nu}+\gamma_{\mu\nu})dx^{\mu}dx^{\nu}
+\gamma_{\mu5}(x,y)dx^{\mu}dy+2(1+\gamma_{55}(x,y))dy^2\rt],
\label{Fluc_Aux_Metric_UC}
\eeqn
where $\xi$ and $\gamma$ represent fluctuations of the auxiliary metric.

Now, with these two perturbed metrics, after calculating the linear fluctuations of the field equation (\ref{EOM3}), the perturbed auxiliary metric is given by
\begin{subequations}\label{Sol_Fluc_Aux_Metric}
\beqn
\gamma_{\mu\nu}&=&h_{\mu\nu}+\fc{1}{Ka^2}(3\fc{a'}{a}\delta\phi-\fc{1}{6}\delta\phi')\eta_{\mu\nu},\\
\gamma_{\mu 5}&=&\fc{1}{Ka^2}\pt_\mu\delta\phi,\\
\gamma_{55}&=&\fc{4}{3Ka^2}\delta\phi',
\eeqn
\end{subequations}
where $\delta\phi$ represents the perturbation of the background scalar field.
Further, with the above perturbed auxiliary metric and the relations (\ref{EOM_EP_1}) and (\ref{EOM_EP_2}), the linear fluctuations of the field equation (\ref{EOM2}) are give as follows.

The $\mu\nu$ components:
\beqn
&&\fc{1}{2}\eta^{\lambda\sigma}\pt_\lambda\pt_\mu{h}_{\nu\sigma}+\fc{1}{2}\eta^{\lambda\sigma}\pt_\lambda\pt_\nu{h}_{\mu\sigma}-\fc{1}{2}\Box^{(4)}{h}_{\mu\nu}-\fc{1}{4}a^2{h}_{\mu\nu}''
-2aa'{h}_{\mu\nu}'-\fc{1}{2}\pt_{\mu}\pt_{\nu}{h}-\fc{7}{12}aa'\eta_{\mu\nu}{h'}\nn\\
&&+\fc{1}{24K}\eta_{\mu\nu}\delta\phi'''+\fc{1}{12K}\fc{1}{a^2}\eta_{\mu\nu}\Box^{(\!4\!)}\delta\phi'\!-\!\fc{1}{3K}\fc{a'}{a^3}\eta_{\mu\nu}\Box^{(\!4\!)}\delta\phi\!-\!\fc{1}{K}\fc{a'}{a^3}\pt_\mu\pt_\nu\delta\phi
+\fc{7}{12K}\fc{a'}{a}\eta_{\mu\nu}\delta\phi''\nn\\
&&-\fc{1}{36K}\fc{a''}{a}\eta_{\mu\nu}\delta\phi'-\fc{13}{36K}\fc{a'^2}{a^2}\eta_{\mu\nu}\delta\phi'-\fc{3}{4K}\fc{a'''}{a}\eta_{\mu\nu}\delta\phi-\fc{25}{4K}\fc{a''a'}{a^2}\eta_{\mu\nu}\delta\phi+\fc{30}{K}\fc{a'^3}{a^3}\eta_{\mu\nu}\delta\phi\nn\\
&&-\fc{3}{bK}\fc{a'}{a}\eta_{\mu\nu}\delta\phi+\fc{1}{6bK}\eta_{\mu\nu}\delta\phi'=0,
\label{Fluc_Eq_1}
\eeqn
where $h=\eta^{\lambda\sigma}h_{\lambda\sigma}$ and $\Box^{(4)}=\eta^{\lambda\sigma}\pt_{\lambda}\pt_{\sigma}$.

The $\mu 5$ component:
\beqn
&&\fc{1}{2}\eta^{\lambda\sigma}\pt_{\lambda}h_{\mu\sigma}'-\fc{1}{2}\pt_{\mu}h'+\fc{1}{4K}\fc{1}{a^2}\pt_\mu\delta\phi''-\fc{1}{3K}\fc{a'}{a^3}\pt_\mu\delta\phi'
-\fc{9}{2K}\fc{a''}{a^3}\pt_\mu\delta\phi+\fc{27}{2K}\fc{a'^2}{a^4}\pt_\mu\delta\phi\nn\\
&&-\fc{1}{bK}\fc{1}{a^2}\pt_\mu\delta\phi=0.
\label{Fluc_Eq_2}
\eeqn

The $55$ component:
\beqn
&&-\fc{1}{2}h''-\fc{5}{3}\fc{a'}{a}h'+\fc{1}{3K}\fc{1}{a^2}\delta\phi'''-\fc{1}{3K}\fc{1}{a^4}\Box^{(4)}\delta\phi'
-\fc{2}{3K}\fc{a'}{a^5}\Box^{(4)}\delta\phi+\fc{10}{3K}\fc{a'^2}{a^4}\delta\phi'-\fc{2}{9K}\fc{a''}{a^3}\delta\phi'\nn\\
&&-\fc{6}{K}\fc{a'''}{a^3}\delta\phi+\fc{34}{K}\fc{a''a'}{a^4}\delta\phi-\fc{12}{K}\fc{a'^3}{a^5}\delta\phi-\fc{8}{3bK}\fc{1}{a^2}\delta\phi'=0.
\label{Fluc_Eq_3}
\eeqn

Moreover, the fluctuation of the scalar field equation (\ref{Scalar_EOM}) is given by
\beqn
\fc{1}{2}\phi'h'+\fc{1}{a^{2}}\Box^{(4)}\delta\phi+\delta\phi''+4\fc{a'}{a}\delta\phi'=\fc{\pt^2V}{\pt\phi^2}\delta\phi.
\label{Fluc_Eq_4}
\eeqn

Further, we consider the transverse-traceless (TT) components $\bar{h}_{\mu\nu}$ of the gravitational perturbations defined by $\bar{h}_{\mu\nu}={P_{\mu\nu}}^{\sigma\rho}h_{\sigma\rho}$, where
$P_{\mu\nu\sigma\rho}=\Pi_{\mu\sigma}\Pi_{\rho\nu}-\fc{1}{3}\Pi_{\mu\nu}\Pi_{\sigma\rho}$ is the TT projection operator for symmetric tensor field with $\Pi_{\mu\nu}=\eta_{\mu\nu}-{\pt_{\mu}\pt_{\nu}}/{\Box^{(4)}}$. And the TT components satisfy the conditions
\beqn
\eta^{\mu\nu}\bar{h}_{\mu\nu}=\eta^{\lambda\nu}\pt_{\lambda}\bar{h}_{\mu\nu}=0.
\eeqn
Since Eqs. (\ref{Fluc_Eq_2}), (\ref{Fluc_Eq_3}) and (\ref{Fluc_Eq_4}) are purely non-TT components and all the TT components are involved in the Eq. (\ref{Fluc_Eq_1}), with the TT projection operator, it is easy to find that the scalar perturbation $\delta\phi$ decouples from the TT tensor perturbations and Eq. (\ref{Fluc_Eq_1}) gives us
\beq
\fc{1}{2}a^2\bar{h}_{\mu\nu}''+4aa'\bar{h}_{\mu\nu}'+\Box^{(4)}\bar{h}_{\mu\nu}=0.
\label{TT_EQ_UC}
\eeq
In order to eliminate the prefactor $a^2$ in the first term, we utilize a coordinate transformation $dy=\fc{a(z)}{\sqrt{2}}dz$ to rewrite the perturbed metric (\ref{Fluc_Metric_UC}), and hence,
in the new coordinate $z$, Eq. (\ref{TT_EQ_UC}) is rewritten as
\beq
\pt_{z,z} \bar{h}_{\mu\nu}+7\fc{\pt_z a}{a}\pt_z \bar{h}_{\mu\nu}+\Box^{(4)}\bar{h}_{\mu\nu}=0.
\label{Fluc_Eq_C}
\eeq
Further, we decompose $\bar{h}_{\mu\nu}$ in the form
\beq
\bar{h}_{\mu\nu}(x,z)=\varepsilon_{\mu\nu}(x)a^{-\fc{7}{2}}(z)\Psi(z)\label{KK_Decomposition},
\eeq
where the factor $a^{-7/2}(z)$ is appended in order to eliminate the first derivative term of $\Psi(z)$, and the mass $m$ of KK excitations is defined by the four-dimensional Klein-Gordon equation
\beq
\Box^{(4)}{\varepsilon_{\mu\nu}(x)}=m^2 \varepsilon_{\mu\nu}(x). \label{4D_Fluctuation}
\eeq
Then a Schr$\ddot{\text{o}}$dinger-like equation is obtained from Eq. (\ref{Fluc_Eq_C})
\beq
-\pt_{z,z}\Psi(z)+U(z)\Psi(z)=m^2\Psi(z),\label{Schrodinger_Eq}
\eeq
with the effective potential $U(z)$ given by
\beq
U(z)=\fc{7}{2}\fc{\pt_{z,z}a}{a}+\fc{35}{4}\fc{(\pt_z a)^2}{a^2}.\label{Effective_Potential}
\eeq
The Hamiltonian in Eq. (\ref{Schrodinger_Eq}) can be factorized as
\beq
H=\lt(\fc{d}{dz}+\fc{7}{2}\fc{\pt_z a}{a}\rt)\lt(-\fc{d}{dz}+\fc{7}{2}\fc{\pt_z a}{a}\rt), \label{Ham_Fac}
\eeq
with the coefficient $7/2$ rather than $3/2$ as usual \cite{Randall1999a,Gremm2000a,DeWolfe2000,Kobayashi2002,Wang2002,Bazeia2004a,Yang2011}. Thus, supersymmetric quantum mechanics ensures that there is no normalizable mode with $m^2<0$ \cite{Csaki2000}. It means that this system is tachyonic free and stable under tensor fluctuations. The KK mass spectrum of the Schr$\ddot{\text{o}}$dinger equation (\ref{Schrodinger_Eq}) determines graviton masses observed on the brane. The zero mode is easily got from (\ref{Schrodinger_Eq}) by setting $m=0$, and it is
\beq
\Psi_0(z)=N_0~a^{7/2}(z),
\eeq
where the normalization parameter $N_0$ is fixed by the normalization condition
$\int\Psi^2(z)dz=\int\Psi^2(y)\fc{dy}{a(y)}=\int N_0^2a^6(y)dy=1$, so $N_0^2=45/[256\sqrt{b}\sqrt{\fc{2}{21\pi}}\Gamma(\fc{9}{4})\Gamma(\fc{13}{4})]\approx{0.35}/{\sqrt{b}}$.
The normalizable zero mode ensures that the massless graviton only propagates along the brane and provides the gravitational fields in the low-energy effective theory.
%%%%%%%%%%%%%%%%%%%%%%%%%%%%%%%%%%%%%%%%%%%%%%%%%%%%%%%%%%%%%%%%
\begin{figure*}[htb]
\begin{center}
\subfigure[$U(z)$]  {\label{Gra_potential}
\includegraphics[width=7cm,height=5cm]{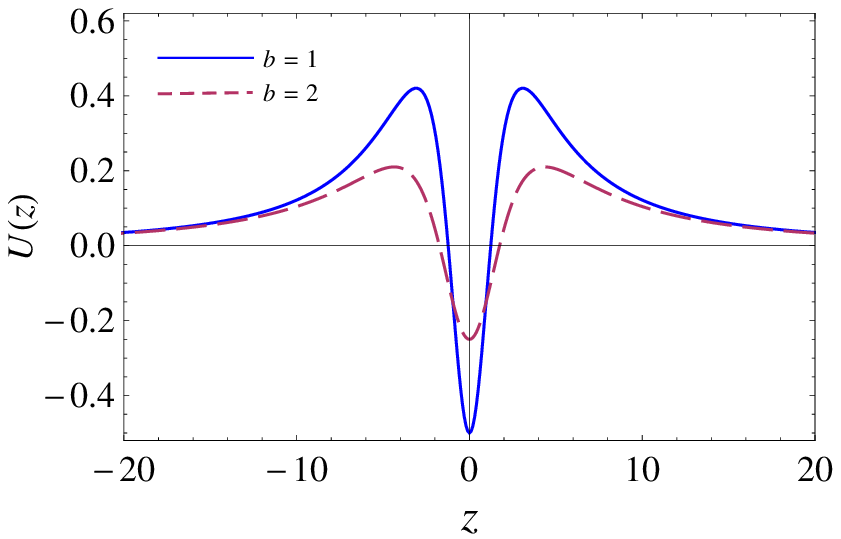}}
\subfigure[$\Psi_0(z)$]  {\label{Zero_mode}
\includegraphics[width=7cm,height=5cm]{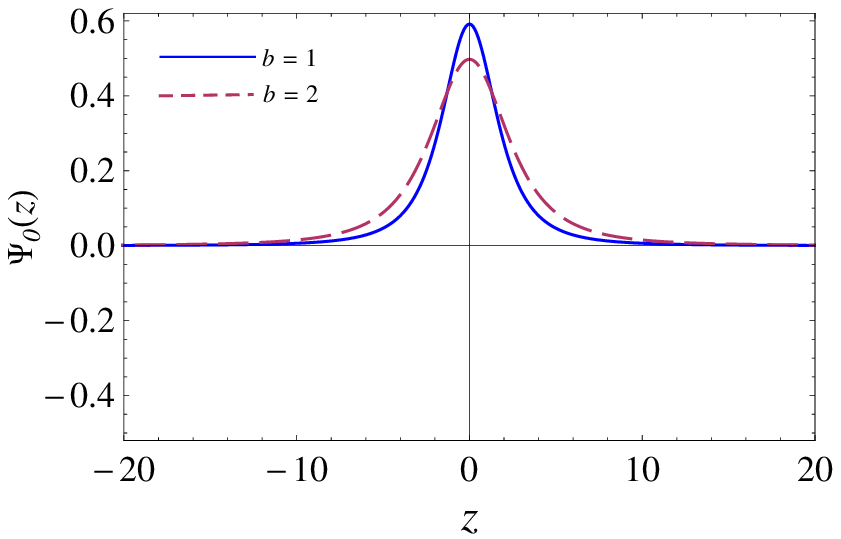}}
\end{center}
\caption{The shapes of the effective $U(z)$ potential and the zero mode $\Psi_0(z)$. The
parameters are set to $ \kappa=1, \lambda=1$.}
\label{Gra_potential_zero_mode}
\end{figure*}
%%%%%%%%%%%%%%%%%%%%%%%%%%%%%%%%%%%%%%%%%%%%%%%%%%%%%%%%%%%%%%%%%%

We numerically plot the effective potential $U(z)$ and the zero mode $\Psi_0(z)$ in Fig. \ref{Gra_potential_zero_mode}.
The Fig. \ref{Zero_mode} shows that the gravitational zero mode localizes at the origin of extra dimension and vanishes as $|y|\rightarrow\infty$. The effective potential is volcano-like. It has a deep well
at the origin and asymptotically vanishes at each side, as shown in Fig. \ref{Gra_potential}. Thus, besides a bound massless mode, there exists a set of continuous massive KK modes $\Psi_m(z)$ starting at $m^2>0$. These massive KK modes are not localized on the brane. Nevertheless, for two barriers existing at each side of the potential well, some massive resonant states could exist. If their lifetimes on the brane are long enough, they can be regarded as quasilocalized massive gravitons.

From the decomposition (\ref{KK_Decomposition}), the wave function of the gravitational zero mode is $\bar{h}^{(0)}_{\mu\nu}(x)=N_0\varepsilon_{\mu\nu}(x)$, and this TT mode is the four-dimensional massless graviton propagating on the brane. Therefore, as in \cite{Randall1999a}, the low-energy effective theory should be provided by including only the massless zero mode in the
fluctuational metric (\ref{Fluc_Metric_UC}), i.e.,
\beqn
d\hat{s}^2=a^2(y)g^{(4)}_{\mu\nu}(x)dx^{\mu}dx^{\nu}+dy^2
=a^2(y)(\eta_{\mu\nu}+\bar{h}^{(0)}_{\mu\nu}(x))dx^{\mu}dx^{\nu}+dy^2.
\eeqn
Since the scalar perturbation decouples from the TT components of the tensor perturbations and is just relevant to non-TT components, from the linear fluctuations of the field equations, the above effective perturbed metric with the TT zero mode just leads to the vanishing of the scalar perturbation $\delta\phi$. And hence, the components of the perturbed auxiliary metric (\ref{Sol_Fluc_Aux_Metric}) are simplified as $\gamma_{\mu\nu}=\bar{h}^{(0)}_{\mu\nu}$ and $ \gamma_{\mu5}=\gamma_{55}=0$. Thus the perturbations of the auxiliary metric are identical to the TT perturbations in the space-time metric although they are multiplied by different warp factors. This intriguing result was also presented in \cite{Escamilla-Rivera2012}, where the authors studied the TT tensor perturbations of a homogeneous and isotropic space-time in EiBI theory. However, when the non-TT part is included, this conclusion does not hold anymore for the nonvanishing $\delta \phi$ in (\ref{Sol_Fluc_Aux_Metric}). Then, with this perturbed auxiliary metric, the EiBI action (\ref{EiBI_action}) gives us the four-dimensional effective gravitational theory on the brane
\beqn
S\supset\fc{1}{{\kappa}b}\int{d^5x\lt[\sqrt{-|\hat{g}_{MN}+b{R}_{MN}|}-\lambda\sqrt{-|\hat{g}_{MN}|}\rt]}
                      \supset\fc{1}{{2\tilde\kappa}}\int{d^4x\sqrt{-|g^{(4)}_{\alpha\beta}|}R^{(4)}},
\label{Low_E_Th}
\eeqn
where $R^{(4)}=g^{(4)\mu\nu}R^{(4)}_{\mu\nu}$ with $R^{(4)}_{\mu\nu}(x)$ constituting by $g^{(4)}_{\mu\nu}(x)$, and $1/\tilde\kappa=\fc{\sqrt{2}\alpha^{\fc{3}{2}}}{\kappa}\int_{-\infty}^{+\infty}{dya^{6}(y)}=\sqrt{2}\alpha^{\fc{3}{2}}/(N_0^2\kappa)\approx{5.09\sqrt{b}}/{\kappa}$. So, we can read off the relation of the effective four-dimensional Planck scale $M_N=(\tilde\kappa)^{-1/2}$ and the fundamental five-dimensional Planck scale $M_{*}=(\kappa)^{-1/3}$, namely, $M^2_{N}=5.09\sqrt{b}M^3_{*}$. An analogous relation can be found in Randall-Sundrum-2 model \cite{Randall1999a}, where $M^2_N=M^3_*/k$ with the parameter $k$ a scale of order the Planck scale $E_{\text{Pl}}\sim10^{19}$GeV. Thus, here we set the parameter $b{\sim}E_{\text{Pl}}^{-2}$ to fix $M_N$ and $M_*$ both of order the Planck scale. The Eq. (\ref{Low_E_Th}) shows that four-dimensional Einstein gravity is indeed recovered on the brane at low-energy level.

\section{Conclusions and Discussion}

In this paper, we have investigated generating a flat thick brane configuration in EiBI gravity by including a scalar field in the bulk. The solution of scalar field is found to be a kink which connects two vacua of the potential and the brane configuration is a domain wall. Further, as discussions in \cite{Liu2011,Cheng2010,Liu2008,Liu2009c}, by introducing a proper Yukawa coupling with the scalar field, Dirac fermions can be localized on the domain wall.

In tensor fluctuations, a Schr$\ddot{\text{o}}$dinger equation is obtained, and furthermore, the Hamiltonian can be factorized. It ensures the stability of the system. While in \cite{Escamilla-Rivera2012}, the tensor modes were found to be linearly unstable in TT perturbations to a four-dimensional homogeneous and isotropic universe, thus the brane-world scenario may be helpful to stabilize the models in this gravity theory.  For the volcano-like potential, there exists just one bound zero mode (massless graviton observed on the brane) and a set of continuous massive modes. The normalized zero mode provides the four-dimensional gravitational fields in the low-energy effective theory. As shown in Eq. (\ref{Low_E_Th}), GR is indeed recovered on the brane, and hence, it ensures the theory does not violate experimental observations.

As $\lambda$ is a dimensionless parameter, $b\sim{E_{\text{Pl}}^{-2}}$ and $\kappa\sim{E_{\text{Pl}}^{-3}}$, the scalar vacuum energy density is fixed as $V_0\sim{E_{\text{Pl}}^{5}}$. On the other hand, when $y\rightarrow\pm\infty$, namely, away from the scalar source, the system approaches AdS vacuum. We notice that the constant $V_0=-\fc{\lambda}{b\kappa}$ in the scalar potential solution (\ref{Scalar_Potential}) cancels out the $\lambda$ term in the EiBI action (\ref{EiBI_action}), thus the effective five-dimensional cosmological constant can be easily read from Eq. (\ref{Vac_Eq1}) as $\Lambda_5=-\fc{3}{2b}$. It is consistent with the statement that the geometry of the bulk is an asymptotically AdS space-time ($R\rightarrow-5/b$). Although the five-dimensional cosmological constant $\Lambda_5$ is huge, the brane is still effectively flat (four-dimensional Minkowski space-time).

The parameter $b$ is set to be small enough, while the Ricci tensor $R_{MN}$ is proportional to $1/b$, thus the expansion (\ref{Smal_Appro_Action}) is invalid, and the EiBI theory deviates from GR. On the other hand, owing to the connection constituted by the auxiliary metric which deviates greatly from the space-time metric when the background scalar included, the potential of the Schr$\ddot{\text{o}}$dinger equation is distinct from the one achieved in the brane model based on GR. And, it just emerges as one apparent feature of the deviation. Since the Schr$\ddot{\text{o}}$dinger equation determines the mass spectrum of KK gravitons, and further, effective Newtonian potential between two particles located on the brane is generated by exchange of the zero mode
and massive KK modes \cite{Randall1999a,Callin2004a}, the corrections of Newtonian potential should be different. The explicit potential correction is left for future works.

\acknowledgments{
This work was supported by the Program for New Century Excellent Talents
in University, the National Natural Science Foundation of China (Grant No. 11075065),
the Doctoral Program Foundation of Institutions of Higher Education of China
(Grant No. 20090211110028), the Huo Ying-Dong Education Foundation of
Chinese Ministry of Education (Grant No. 121106),  and the Fundamental Research Funds for the Central Universities (Grant No. lzujbky-2012-k30).}

\end{document}